# Topologically protected entangled photonic states

Michelle Wang[1], Cooper Doyle[1], Bryn Bell[1,2], Matthew J. Collins[3], Eric Magi[1,4], Benjamin J. Eggleton[1,4], Mordechai Segev[5], and Andrea Blanco-Redondo[1,4,*]

1. Institute of Photonics and Optical Science (IPOS), School of Physics, The University of Sydney, Sydney, New South Wales 2006, Australia.
2. Clarendon Laboratory, Department of Physics, University of Oxford, Oxford, UK.
3. MQ Photonics Research Centre, Department of Physics and Astronomy, Macquarie University, Sydney NSW 2109, Australia.
4. The University of Sydney Nano Institute (Sydney Nano), New South Wales 2006, Australia.
5. Physics Department and Solid State Institute, Technion-Israel Institute of Technology, Haifa 32000, Israel

* andrea.blancoredondo@sydney.edu.au

**Entangled multiphoton states lie at the heart of quantum information, computing, and communications. In recent years, topology has risen as a new avenue to robustly transport quantum states in the presence of fabrication defects, disorder and other noise sources. Whereas topological protection of single photons and correlated photons has been recently demonstrated experimentally, the observation of topologically protected entangled states has thus far remained elusive. Here, we experimentally demonstrate the topological protection of spatially-entangled biphoton states. We observe robustness in crucial features of the topological biphoton correlation map in the presence of deliberately introduced disorder in the silicon nanophotonic structure, in contrast with the lack of robustness in nontopological structures. The topological protection is shown to ensure the coherent propagation of the entangled topological modes, which may lead to robust propagation of quantum information in disordered systems.**

The discovery of topological insulators has stimulated the design of topological systems across many platforms beyond condensed matter, including electromagnetism [1]-[9], ultracold atoms [10],[11] and phonons [12]-[14]. In all these systems, the unique topology of the wave functions in the band structure gives rise to protected edge modes which are inherently robust against disorder.

Analogous to the original condensed matter systems, topological photonic systems provide an avenue for the robust transport of light. Classical light experiments have demonstrated topologically protected edge states in both the microwave [1],[6],[9], and optical regimes [2]-[5],[7],[8] with potential applications in robust optical delay lines [15], lasing in topological defects [16]-[18], and eventually topological insulator lasers [19]-[21]. Recently, progress in topological photonics has extended into the quantum regime [22]-[32]. Photons remain intrinsically well-isolated from the environment, even at room temperatures, but scattering loss, absorption, and phase errors are still hindering the scalability of photonic quantum computing and communication systems. Single-photon experiments in free space have demonstrated topological protection in discrete-time [22],[23],[27],[29]. However, the real advantage of using topological settings for quantum optics experiments should be in integrated platforms, where the topological protection can provide robustness against defects, inhomogeneities and disorder [24],[25]. In 2018, the first topological experiments with single photons in integrated platforms were reported, with the demonstration of an integrated photonics interface between a quantum emitter and counterpropagating single-photon edge states [28] and of high-visibility quantum interference of topological single-photon states [31], important steps for linear optics-based quantum computation. The photon number distribution of single photons shows lower variance than classical light, hence such experiments could have an impact in achieving lower noise sources. However, the propagation and manipulation of multiphoton states lie at the heart of quantum information systems and thus a platform which can protect multiphoton states is of great interest. A recent experiment has shown lattice topology giving rise to spectrally robust generation of photon pairs [30]. Independently, our team has demonstrated topological protection of the biphoton correlation in a CMOS-compatible lattice of coupled silicon nanowires [32]. Nonetheless, until now, the experimental demonstration of topological protection of entangled states remained a challenge.

Here, we present the first experiments demonstrating topological protection of entangled biphoton states. We fabricate and measure a topological nanophotonic system based on the Su-Schrieffer-Heeger (SSH) model [33] with two spatially-entangled biphoton states, as theoretically suggested in [32]. The silicon nanowires are arranged in a one-dimensional lattice with staggered short and long gaps which determine the coupling constants between adjacent nanowires [8],[32]. The introduction of long-long defects into the gaps results in topological interfaces between two mirror images of the SSH lattice and yields topological modes localized at the interfaces. Our lattice contains two of these long-long defects. We generate multiphoton spatial entanglement between these modes, using on-chip generated correlated photon pairs, and we experimentally demonstrate the robustness of the spatially entangled biphoton correlation map against disorder in the position of the silicon nanowires.

**METHODS**

Figure 1A presents a scanning electron microscope (SEM) image of one of the fabricated lattices with two long-long topological defects separated by 5 dimers (10 nanowires) to avoid coupling between the resulting two topological defect modes. The complete lattice contains 202 silicon nanowires with length $L = 381$ μm, height $h = 220$ nm and width $w = 450$ nm on a silica substrate. The gaps separating the waveguides alternate between short gap $g_s = 173$ nm and long gap $g_l = 307$ nm, in resemblance to the SSH model. The two center waveguides of the long-long defects are shaded in yellow.

To faciliate comparison, we fabricate and measure equivalent arrays with two topologically-trivial defects, also separated by 5 dimers to avoid coupling between the two resulting trivial modes. These arrays are comprised of 202 equidistant nanowires with two wider waveguides that create regions with a higher effective index of refraction within the lattice, giving rise to two trivially localized defect modes. Figure 1B shows an SEM image of one such lattice, with the wider waveguides shaded in yellow. We keep the nanowire dimensions the same but choose a gap $g = 100$ nm between all nanowires and a width $w_c = 465$ nm for the two center waveguides to achieve the same modal confinement as the topological system.

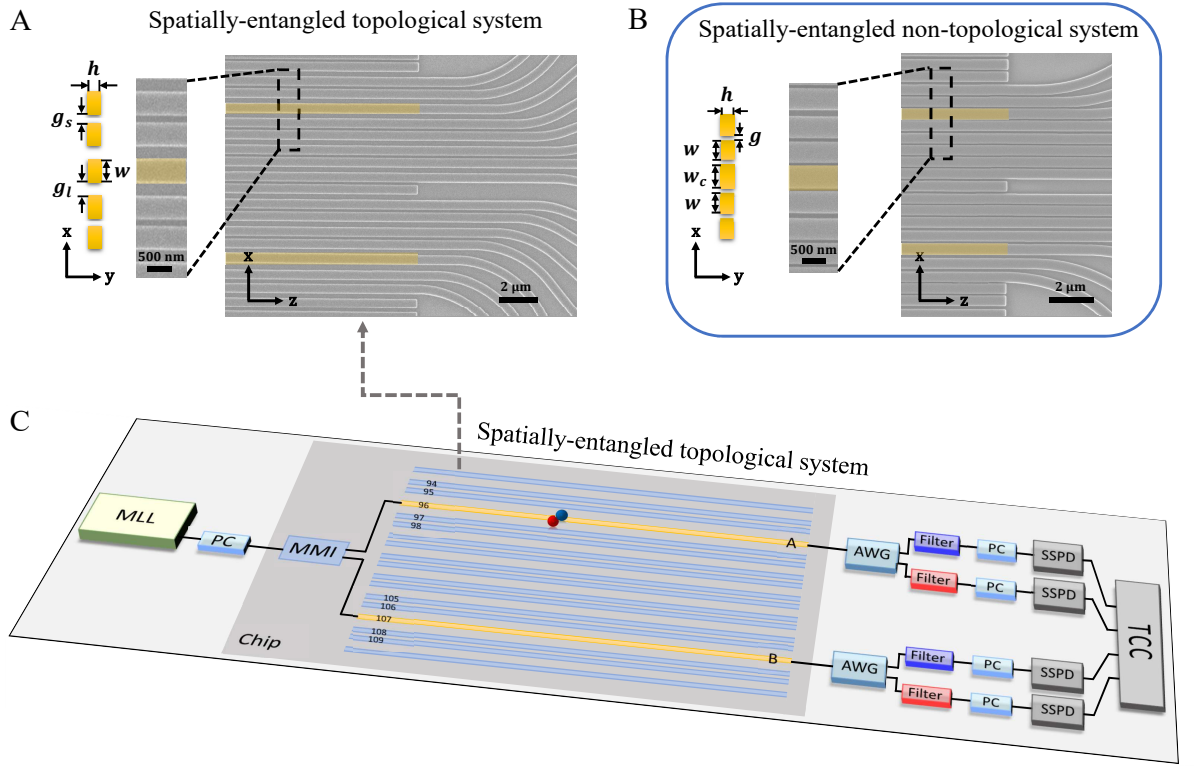

**Fig. 1:** Experimental setup of our nanophotonic platform. (**A**) SEM image of the center waveguides of a fabricated lattice with 2 topological defects. The center waveguide of each long-long defect is highlighted in yellow. (**B**) SEM image of center waveguides of non-topological system with two trivial defects, where localization is provided by index-guiding. The two wider waveguides are highlighted in yellow. (**C**) Schematic of the experimental setup. Mode-locked laser (MLL), polarization controllers (PCs), on-chip multimode interferometer (MMI), lattice of coupled silicon waveguides with two long-long defects supporting topological modes localized around waveguides 96 and 107, arrayed waveguide gratings (AWGs), tunable filters, superconducting nanowire single-photon detectors (SSPDs), and time correlation circuit (TCC).

We generate the entangled biphoton states via spontaneous four wave mixing (SFWM) within the photonic chip itself. We launch picosecond pulses at 1550 nm emitted by a mode-locked laser at the input of the chip, by splitting into two arms of equal intensity by a multimode interferometer (MMI), as shown in Fig. 1C. Each arm is then directed into the center waveguide of each of the two long-long (wider waveguide) defects to excite the two topological (trivial) modes of the lattice. As the short pulses propagate through the silicon nanowires, the high peak power and spatial confinement of the light results in the probabilistic generation of correlated signal and idler photon pairs over a broad frequency range via SFWM. These frequency-correlated photon pairs generate spatial entanglement as they couple strongly to the two topological (trivial) modes of the lattice. At the output of the lattice, we filter signal and idler photons at 1545 nm and 1555 nm respectively, which satisfies energy conservation, and individually detect them using superconducting nanowire single-photon detectors (SSPDs). Matching arrival times of the signal and idler photon pairs are measured through a time correlation circuit (TCC) and thus, we map out the spatial profile of the entangled biphoton states.

## RESULTS

### *Entangled biphoton correlation maps*

We measured the biphoton correlation map at the output of two topological lattices with different levels of deliberately-introduced disorder in the waveguide positions, which is manifested as disorder in the coupling constants. Fig. 2A (Fig. 2B) shows the biphoton correlation counts over a 30-minute measurement for a lattice with two topological modes and disorder of 40% (60%) in the coupling constants, which we implement by varying the gaps between waveguides within the range of a few tens of nanometers. Notice that this disorder, although very significant, preserves the chiral symmetry of the structure, and therefore topological protection is expected. For each lattice, the output from ten waveguides were collected – five around each defect, including the two waveguides at the center of the topological long-long defects (labelled as 96 and 107) and two waveguides on either side of those two.

To emphasize the special nature of these topological lattices we measured the biphoton correlation maps at the outputs of two trivial lattices also with increasing levels of deliberately-introduced disorder in the positions of the waveguides (Fig. 2C and 2D). Since the gaps in the trivial lattice are much smaller, we introduced lower levels of disorder in the coupling constant, 0% and 40% respectively. This is to avoid breaking the design rules provided by the nanofabricators, as implementing the levels of disorder used in the topological lattices would lead to instances with excessively small gaps between waveguides that would most likely not be reproduced accurately in the fabricated structure.

To compare with theory, we performed propagation simulations on each measured lattice using the nearest-neighbor tight-binding model (Fig. 2E-2H), and observe agreement in the key features between the experimental measurements and simulated correlation maps. A formal description of the Hamiltonians describing the system and the biphoton dynamics is given in the Supplementary Material of [32].

In both topological lattices, we observe distinct local peaks at waveguides 96 and 107, which show the strong localization of the topological modes to the center of the long-long defects. This corresponds to when the signal and idler photons are both in the center waveguides. Other notable counts include the signal and idler photons both in the next-adjacent waveguides, and the case when the signal (idler) is in center and the idler (signal) is in the next-adjacent. Most importantly, we clearly observe spatial entanglement between two modes that show the spatial signature of the topological modes in the SSH model: zero counts at every other element of the biphoton correlation map. The preservation of this feature between the two lattices with differing levels of disorder in the waveguide positions (Fig. 2A and 2B) is strong evidence of the topological nature of this entangled state. The asymmetry between the two-peaks observed in the measured biphoton correlation maps does not contradict this claim and it is indeed expected in the presence of disorder. As discussed in [32], the topological protection in this system guarantees the preservation of two important features of each of the topological modes: the zero-amplitude in the odd elements of the correlation map and the propagation constant of each of the entangled modes. This is further supported by the simulations in Fig. 2E and 2F, which account for the deliberately introduced disorder in the position of the waveguides and, whereas they clearly show the preservation of the topological features of the entangled state, they also show asymmetry between the two peaks.

For comparison, we now consider the two trivial lattices with 0% and 40% disorder in the coupling constants, whose measured biphoton correlation maps are shown in Fig. 2C and 2D. The biphoton correlations at the output also show distinct peaks at the two center waveguides, highlighting the entanglement between the two trivial modes. However, since these are merely due to index-guiding, the measured biphoton correlations demonstrate no preserved features and robust transport cannot be achieved with this topologically-trivial system. The corresponding simulations in Fig. 2G and 2H reflect this lack of protection in the presence of disorder.

To fully understand every feature in the measured biphoton correlations, it is crucial to appreciate that the deliberately introduced disorder in the gaps at the design phase is not the only disorder present in the system. In this one-dimensional SSH lattice, the entangled topological system is only robust against disorder which preserves the chiral symmetry. In an SSH lattice such as ours, variations in the widths of the waveguides do not preserve chiral symmetry. The variations in the waveguide widths lead to changes in the propagation constant of each individual waveguide. This in turn alters the on-diagonal elements of the Hamiltonian describing the system, which results in the two anomalous non-zero counts in Fig. 2B (inset) where a wider waveguide 108 conduces to some index localization. The discrepancy between the measured and simulated trivial biphoton correlation in the absence of disorder, shown in Fig. 2C and 2G respectively, are also explained from this undesired on-diagonal disorder perspective. These variations in the widths due to fabrication imperfections are very hard to characterize accurately, since they appear at random positions and are often below the resolution of the SEM. Therefore, we have not included this type of disorder in the simulations shown in Fig. 2E-2H. Finally, the dramatic difference in the number of correlated counts in the two center waveguides of Fig. 2B is likely due to the uneven splitting of light by the MMI, also due to fabrication imperfections. As we discuss in the next section, the impact of these types of disorder can be reduced by optimizing the design to be more robust to them.

A formal description of the entanglement in these systems with two defects and a common pump is given in the Supplementary Material of [32] using the Schmidt decomposition. Here, we have computed a lower bound on the Schmidt number for the four measured biphoton correlation maps shown in Fig. 2A-2D and obtained 2.1512, 1.5088, 2.7714 and 2.7702 respectively. The fact that the Schmidt number is always well above one, highlights the strong spatial entanglement present in all our measurements.

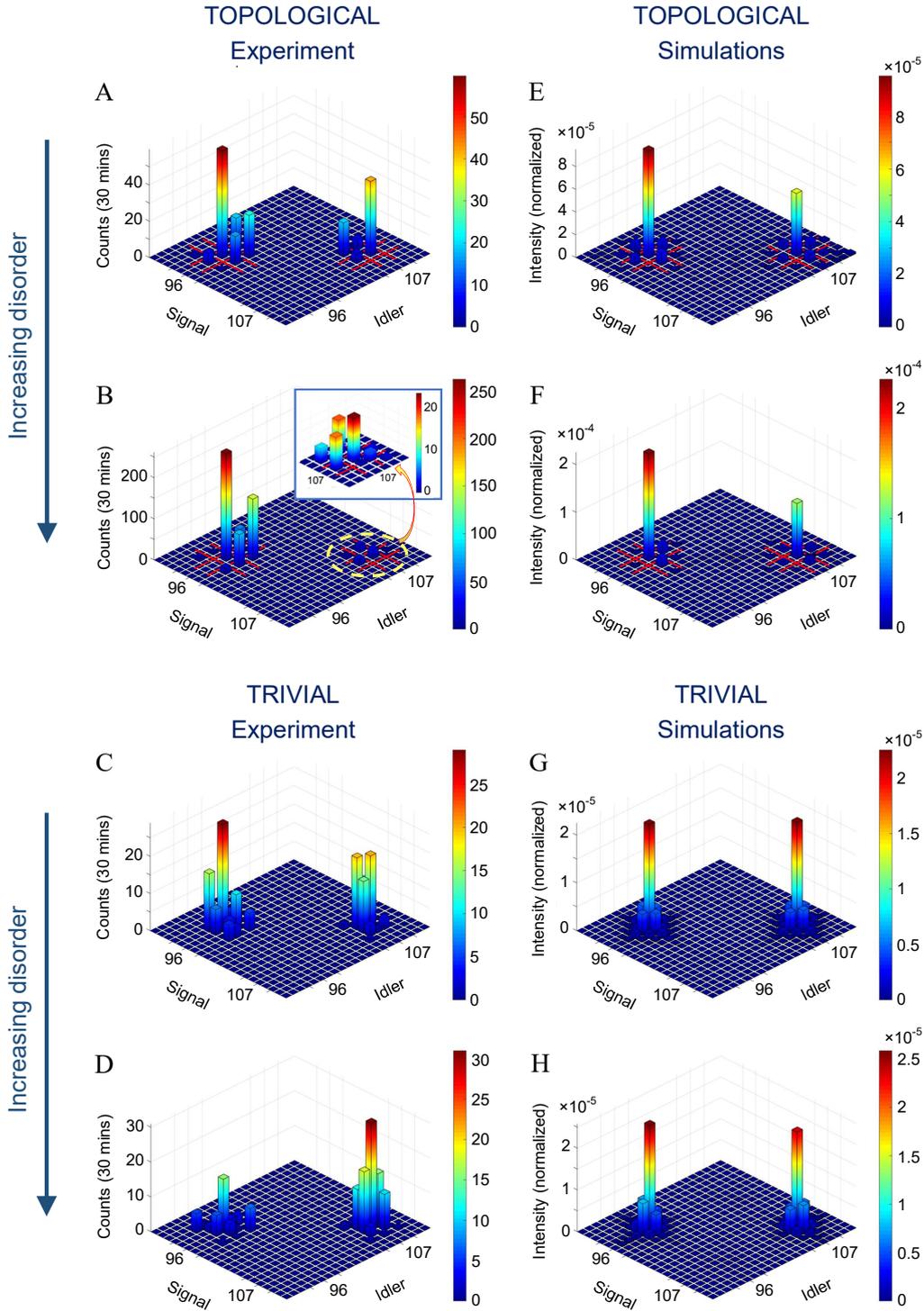

**Fig. 2:** Biphoton states of topological and trivial lattices. (A and B) Measured biphoton correlations for topological lattices with two long-long defects and deliberately-introduced random disorder in the waveguide positions. Every second element has zero counts – a feature of the topological modes. (C and D) Measured biphoton correlations for trivial lattices with random disorder. No features in the correlated counts are preserved. (E to H) Corresponding simulations for the topological and trivial lattices with the disorder included.

## *Propagation Constant*

Let us now consider the other preserved quantity of the topological protection in our system – the fixed propagation constants (analogous to energy in correlated electron systems) of the two defect modes – which relates to phase preservation and is thus of interest for quantum computing and information processing applications. Consider the weak pump regime, where the probability of generating a pair of photons in both topological defect modes simultaneously via SFWM is negligible. The biphoton outputs can then be described as an entangled N00N state

$$|\psi\rangle = \frac{|A\rangle_s|A\rangle_i + e^{2i\phi}|B\rangle_s|B\rangle_i}{\sqrt{2}}$$

where s and i refer to signal and idler photons respectively, and $\phi$ is the phase difference between the two biphoton outputs A and B. This phase difference could arise due to a difference in the propagation constant between the two defect modes, which would yield an accumulated phase difference after a given propagation distance. We can characterize $\phi$ by taking the outputs from the two defects to interfere through a 50:50 beam-splitter. The output after the interferometer can be expressed as a linear superposition of bunched states $|\psi_{bunch}\rangle = \frac{i}{\sqrt{2}}(|B\rangle_s|B\rangle_i - |A\rangle_s|A\rangle_i)$ where the signal and idler appear together, and split states $|\psi_{split}\rangle = \frac{i}{\sqrt{2}}(|A\rangle_s|B\rangle_i + |B\rangle_s|A\rangle_i)$ where the signal and idler appear in different outputs [34],

$$|\psi_{out}\rangle = \cos\phi\,|\psi_{split}\rangle + \sin\phi\,|\psi_{bunch}\rangle.$$

We define the visibility of the system as the normalized difference between the intensities of the split and bunched peaks

$$\text{Visibility} = \frac{\cos^2\phi - \sin^2\phi}{\cos^2\phi + \sin^2\phi} = \cos 2\phi.$$

Note that for two defect modes with the same output phase, i.e. $\phi = 0$, purely split states are obtained, akin to a reverse Hong-Ou-Mandel interferometry process, and the visibility of the system is 1.

We simulate the interference of the output defect modes for the fabricated topological and trivial lattices of Section 3.1, including their specifically designed disorders in the position of the waveguides. Figure 3 shows the biphoton correlation maps after interferometry and their calculated visibilities. As our team discussed in [32], the absence of the two bunch peaks after the beam splitter – out of the four potential peaks: two bunch and two split peaks – is the signature of high visibility quantum interference of the biphoton states. For the topological lattices (Fig. 3A and 3B), we observe that the visibilities remain close to 1, despite the increasing disorder in the positions of the waveguides, highlighting the topological protection of the entangled state. The fact that the visibility is not exactly 1 for the topological lattices is due to the difference in the intensities of the outputs A and B, and in fact, the relative propagation constant of the two topological modes $\Delta k_z = |k_{z,A} - k_{z,B}|$ remains at zero for both lattices.

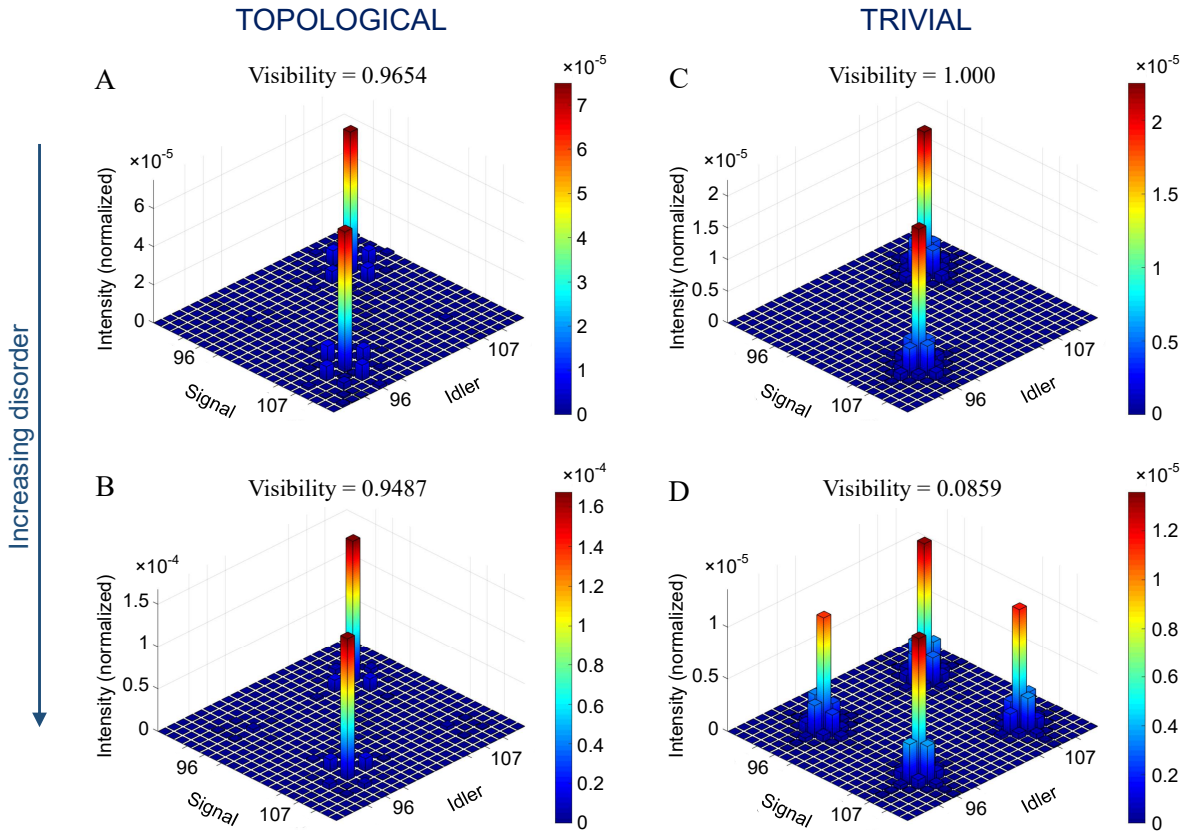

**Fig. 3:** Visibility and biphoton correlation map after interferometry. (A and B) Simulations for fabricated topological lattices from Fig. 2A and 2B respectively. (C and D) Simulations for fabricated trivial lattices from Fig. 2C and 2D respectively.

The trivial lattice with no introduced disorder (Fig. 3C) similarly produces a visibility of 1. However, in the presence of disorder in the position of the waveguides (Fig. 3D), both bunch and split peaks are distinctly present, and the visibility drops dramatically, showing that the output phases of the defect modes are unequal. A direct calculation of the propagation constant mismatch between the two trivial modes from the Hamiltonian of the disordered system measured in Fig. 2D gives $\Delta k_z = 1.83 \times 10^4$ m-1, which results in a large accumulated phase difference of $\phi = 6.96$ radians after propagating through the length of the lattice. Furthermore, this phase difference of the entangled state takes on vastly different values over multiple realizations of the same amount of random disorder. To demonstrate this, we run simulations with different iterations of the same amount of random disorder introduced into the coupling constants (positions of the waveguides) which constitute the off-diagonal elements in the system Hamiltonian. Figure 4A plots the mean absolute deviation of the visibility away from the ideal value of 1 over 100 iterations for each different level of disorder. In the topological lattices (blue), we see that the deviation in the visibility remains close to zero, i.e. the visibility stays close to 1, as disorder is introduced. It increases slightly mostly due to the increasing likelihood of imbalance in the intensities of the peaks caused when one topological defect mode becomes less confined than the other due to the disorder. Indeed, Fig. 4B shows that the mean difference in the propagation constants of the topological defect modes remains effectively at zero. This is strikingly different from the trivial case (Fig. 4A, red) where the mean deviation sharply jumps up as soon as disorder is present. The mean visibility falls close to zero as a result of the phase difference being completely randomized over a wide range (Fig. 4B, red).

For completion, in the Supplementary Material, we characterized the system with respect to a type of disorder that does not preserve the chiral symmetry: disorder in the widths. We verified that the topology cannot protect the entanglement against this type of disorder. However, in our experiment, strong signatures of topological protection in the spatial features of the entangled biphoton states were nonetheless observed also under disorder in the waveguide widths. This allows us to conclude that, even if disorder in the widths is expected from fabrication tolerances, this platform still serves well as a playground for experiments investigating the effect of topology on multiphoton entangled states.

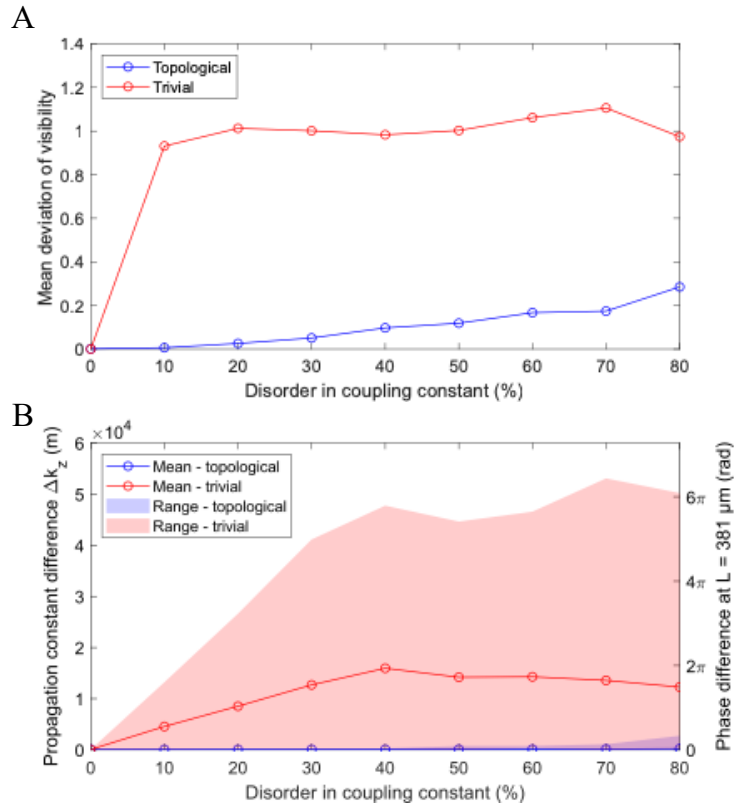

**Fig. 4:** The effect of increasing levels of off-diagonal disorder on topological (blue) and trivial (red) lattices. (A) Mean absolute deviation from visibility = 1 for increasing values of disorder in the coupling constants (manifesting in the waveguide positions). (B) Mean and range of difference in propagation constants of the two defect modes $\Delta k_z$ (left) and the corresponding phase difference at the output of the lattice (right) for increasing disorder in the coupling constants.

**CONCLUSION**

In this article, we presented experimental evidence of biphoton entanglement between two topological modes in a silicon photonics platform. We demonstrated the protection of crucial spatial features of these entangled biphoton states against disorder that preserves the chiral symmetry of the system. In particular, we have shown that the system is robust against fabrication disorder in the positions of the waveguides. Associated with the protection of the spatial features is the other signature of topological protection in SSH systems: the robustness in the propagation constants. We have shown, using simulations, that this signature should manifest in preserving the coherence between topological modes, and therefore in preserving the entanglement. The next step is to measure this second signature experimentally using quantum interference, in line with the experiments in [31].

Altogether, the experiments and simulations presented here provide an experimental proof of concept that topology can indeed protected entanglement against disorder and defects. In our specific photonic platform, being one-dimensional, the protection is against disorder in the positions of the waveguides. However, in photonic platforms of higher dimensionality – such as the helical waveguides [2] and the asymmetrically coupled resonators [3] – the topological protection is expected to be much more general [23], including against disorder in the waveguides width and depth, and also in the coupling between adjacent waveguides. We expect that the robustness of the biphoton correlated map demonstrated here will spark future investigations aiming to encode and manipulate information in the form of qubits on various kinds of topological nanophotonics platforms. In this respect, the experiments presented here constitute the first experimental proofs of concept that topology can indeed provide robustness to entangled photonic states.

## ACKNOWLEDGEMENTS

This work is the result of a close collaboration between the University of Sydney and the Technion. The authors are grateful for a joint grant by the Technion Society of Australia (NSW) together with the NSW Department of Industry. The work was also funded by the Professor Harry Messel Research Fellowship of the School of Physics of the University of Sydney and the Centre of Excellence CUDOS (CE110001018) and Laureate Fellowship (FL120100029) schemes of the Australian Research Council (ARC).

**SUPPLEMENTARY MATERIAL**

In the main text we provided the first experimental demonstration of topological protection of entangled photonic states. The particular platform used to provide this proof-of-concept is a bipartite lattice of silicon nanowires with alternate coupling strengths and two topological defects. This nanophotonics platform implements an SSH Hamiltonian, which is robust against off-diagonal disorder, or, in other words, disorder that preserves the chiral symmetry of the system. We experimentally showed topological robustness of the spatially entangled biphoton states against such type of disorder in the main text.

Here, for the sake of completeness, we consider the effect of disorder in the widths of the waveguides, which affects the on-diagonal elements of the system Hamiltonian and which we know from our measurements that is present in our system. Topological protection in the particular one-dimensional platform does not extend to this type of disorder which breaks the chiral symmetry of the system and thus, the disorder affects both the topological and trivial cases similarly (Fig. S1A and S1B). We have considered a fairly small range of disorder in the widths for our simulations, 0-0.22% of on-diagonal disorder, which corresponds to variations in the width of the waveguides of around ±1 nm. A SEM analysis of several fabricated structures, combined with the e-beam tolerances given by the fabricators, indicate this is a realistic range. Note the high sensitivity of both topological and trivial systems to this type of disorder, with even small deviations in the waveguide widths generating large deviations from the ideal visibility of 1.

Looking towards future applications of this type of structures in quantum information systems, this issue certainly poses a nanofabrication challenge, as the accuracy in the waveguide widths is crucial to preserve topological protection. However, we have verified that at least half of our fabricated topological lattices showed strong evidence of topological protection, indicating that current e-beam fabrication tolerances are not far to meeting the level of accuracy required. Further, this problem can be largely alleviated by using wider waveguides in future designs, given that the same absolute variation in the widths would give rise to much smaller changes in the on-diagonal terms of the Hamiltonian, i.e. in the propagation constant of each of the individual waveguides.

Importantly, the experiments presented in the main text are a proof of concept that topology can protect entangled photonic states. We expect that our results will inspire subsequent experiments aiming for topological protection of the entanglement in a variety of other platforms, including bidimensional platforms that are robust against disorder in the waveguide widths.

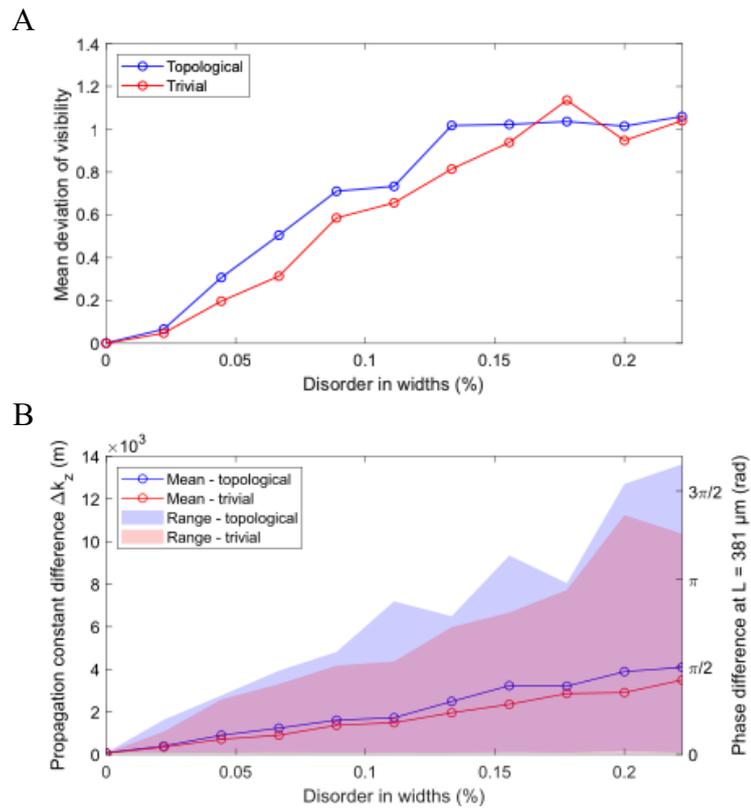

**Fig. S1:** The effect of increasing levels of on-diagonal disorder on topological (blue) and trivial (red) lattices. (A) Mean absolute deviation from visibility = 1 for increasing values of disorder in the propagation constant of the individual waveguides (manifesting in the widths of the waveguides). (B) Mean and range of difference in propagation constants of the two defect modes $\Delta k_z$ (left) and the corresponding phase difference at the output of the lattice (right) for increasing disorder in the waveguide widths. Means and ranges for all cases are calculated over 100 iterations of each level of disorder